\documentclass[conference]{IEEEtran}
\IEEEoverridecommandlockouts
\usepackage{cite}
\usepackage{amsmath,amssymb,amsfonts}
\usepackage{algorithmic}
\usepackage{graphicx}
\usepackage{textcomp}
\usepackage{xcolor}
\usepackage{relsize}
\usepackage{subcaption}
\usepackage{tabularx}
\usepackage{multirow}
\usepackage{cite}
\usepackage{mathtools}
\usepackage{algorithm,algorithmic}

\DeclareUnicodeCharacter{200E}{}
\begin{document}

\title{ Quantifying Uncertainty with Probabilistic Machine Learning Modeling in  Wireless Sensing }

\author{\IEEEauthorblockN{Amit Kachroo}
\IEEEauthorblockA{\textit{Amazon Lab126} \\
Sunnyvale, California, USA  \\
amkachro@amazon.com}
\and
\IEEEauthorblockN{Sai Prashanth Chinnapalli}
\IEEEauthorblockA{\textit{Amazon Lab126} \\
	Sunnyvale, California, USA  \\
	saic@amazon.com}
}

\maketitle

\begin{abstract}
The application of machine learning (ML) techniques in wireless communication domain has seen a tremendous growth over the years especially in the wireless sensing domain. However,   the questions surrounding the ML model's inference reliability, and uncertainty associated with its predictions are never answered or communicated properly. This itself raises a lot of questions on the transparency  of these ML systems. Developing ML systems with probabilistic modeling  can solve this problem easily, where one can quantify uncertainty whether it is arising from the data (irreducible error or aleotoric uncertainty) or from the model itself (reducible or epistemic uncertainty).  This paper describes the idea behind these types of uncertainty quantification in detail and uses a real example of WiFi channel state information (CSI) based sensing for motion/no-motion cases  to demonstrate the uncertainty modeling. This work will serve as  a template to model uncertainty in predictions not only for WiFi sensing but for most  wireless sensing applications ranging from WiFi to millimeter wave radar based sensing that utilizes  AI/ML models. 
	

\end{abstract}


\section{Introduction}
The application of artificial intelligence (AI)/ machine learning (ML) algorithms in wireless sensing applications is seeing an immense growth over the years, and in-fact these models are now embedded into real-world products and features. The main advantages of utilizing AI/ML techniques over conventional principle techniques is the reduced computation complexity, increased energy efficiency, and better optimal solutions\cite{warden2019tinyml}. However, one of the  biggest challenges associated with these AI/ML models is in  its inference reliability or put simply, how confident is the model in its predictions. This unfortunately has not been clearly understood or quantified in the domain of AI/ML in wireless sensing area.  Recently, there has been a lot of research  on this topic \cite{kompa2021second,ovadia2019can,bhatt2021uncertainty}, but these are mostly limited to medical or computer vision field.  This paper will therefore present the science behind the uncertainty modeling, which will help to answer the model reliability. We will dwell into more details about this method with a real example of WiFi channel state information (CSI) based sensing for motion/no-motion detection application.

Generally, uncertainty is classified into two broad categories, aleotoric and epistemic uncertainty\cite{gal,kendall2017uncertainties,hullermeier2021aleatoric}.  Aleotoric derives its name from the Latin word ``alea"  that means the roll of a dice and Epistemic derives its name from the Greek word ``episteme", which can be roughly translated as knowledge. Therefore, aleotoric uncertainty is the  internal randomness of a phenomena  and epistemic is presumed to derive from the lack of knowledge regarding the phenomena. In wireless sensing applications, this is a very common phenomenon where a AI/ML model trained on a certain set of home environments  performs in a very uncertain way when tested on a  different home environment.  The main reason of the failure of these ML models  is because of the model not only learns the features extracted from the radio frequency (RF) signals but also learns the other information from environment, which is not desired in  RF sensing. This extra information is due to the fact that the RF signals are highly dependent on the  scattering conditions  such as   reflections/ diffraction's from the objects in the environment (walls, furniture, etc.), and also on the  position and distance of different objects or human/pets  from the radio device. Therefore, it becomes a necessity for  AI/ML model developed for wireless sensing to communicate its reliability or uncertainty associated with its predictions.

In this work, we will discuss the method to include uncertainty (aleotoric and epistemic) into the ML model, and discuss pros and cons of this approach in detail. In summary,  the main contributions of the paper are,
\begin{itemize}
	\item Understanding different types of uncertainty associated with a AI/ML model.
	\item Modeling uncertainty in a AI/Ml model  with a real life example of WiFi sensing.
	\item Discussions on the results highlighting the need of incorporating uncertainty in wireless sensing applications. 
\end{itemize}
This paper is organized as follows, Section \ref{UncertaintyType} discusses the different types of uncertainty in detail for  AI/ML models, Section \ref{wifiSensing} presents the details of the WiFi CSI based motion/no-motion detection application. In Section \ref{QuantResults}, we will discuss the model and the results of uncertainty quantification for our example  in detail and finally, the conclusion with future work are drawn in Section \ref{ConclusionFuture}. 

\section{Uncertainty in AI/ML models} \label{UncertaintyType}

To start with, the term uncertainty actually in itself means  lack of knowledge to a particular outcome. In AI/ML domain, this can be attributed to either  data itself (measurement noise, or wrong labeling) or to  the  model (model parameters) or lack of  training data.  This is broadly classified as aleotoric and epistemic uncertainty.  
\begin{itemize}
	\item \textit{Aleotoric  or indirect uncertainty}- This type of uncertainty arises from the unaccounted factors, such as environment settings, noise in the input data, or bad input feature selections. It is also known as an irreducible error and can't be remediated with more data. One of the solution to overcome such uncertainty is to make sure that the data collection strategy is carefully designed and  the measurement environment is constrained so that the effect of environment or any external factors  is minimized. Also careful feature selection that represents the phenomena or application  is of utmost importance to avoid such uncertainty.
	\item \textit{Epistemic or direct uncertainty}- This type of uncertainty usually arises from the lack of knowledge about the model or data. One example can be over-generalization, where the ML model is very complex as compared to the amount of data it is trained on and thereby overgeneralizes  on a test dataset. This type of uncertainty can be overcomed by  collecting more data or experimenting with different ML model architectures or by changing/tweaking model parameters.  Since, this uncertainty is caused inherently by the model/data, therefore it can be easily reduced by more data or by different model architecture.
\end{itemize}

Epistemic uncertainty is also used to detect dataset shifts (test data has different distribution than training), or adversarial inputs. Modeling epistemic uncertainty is challenging than modeling the aleotoric one. The later one is incorporated in the model loss function while the epistemic is highly dependent on the model itself and may vary from one model architecture to other.

\subsection{Modeling Aleotoric and Epistemic Uncertainty}

Given a dataset, $\mathbb{D}=\{ X_i,y_i\}, \ i\in\{1,\dots,n\}$, where $X_i$  is the $i^{th}$ input, $y_i$ is the $i^{th}$ output and $n$ is the total number of input examples in the dataset, the ML model can be then described as a function, $\hat{f}:X_i \mapsto \hat{y_i} $, or
\begin{equation}
	\hat{y_i}= \hat{f}(X_i) 
\end{equation}
and lets assume the original data generating process can be given by a function $f:X_i \mapsto y_i$, such that  $y_i= f(X_i)+ \epsilon_i$, where $\epsilon_i$ represents the irreducible error caused by measurement errors during data collection or by wrong labeling in the training data or bad input feature selection. Thus, the mean square error (MSE) between the actual labels and predicted labels from the model will be given as,

\begin{equation}\label{berror}
	\begin{split}
	E(y_i-\hat{y_i})^2 &= E(f(X_i) + \epsilon_i -\hat{f}(X_i)),\\
	&= \underbrace{[f(X_i)-\hat{f}(X_i)]^2}_{\text{reducible error}} + \underbrace{\text{Var}(\epsilon_i) .}_{\text{irreducible error}}
\end{split}
\end{equation}

The first part in \eqref{berror} is model dependent and therefore represents epistemic uncertainty while the second term (variance of $\epsilon_i$)  is the irreducible or aleotoric uncertainty. This variance of $\epsilon_i$  is also known as the Bayes error, which actually is the lowest possible prediction error than can be achieved with any model. In literature, $\epsilon$ is generally modeled as  an independent and identically distributed (i.i.d) Normal distribution, $\epsilon_i\sim\mathcal{N}(\mu_i,\sigma_i)$.  To incorporate the aleotoric uncertainty in a  AI/ML model,  the final  layer  can be therefore  replaced with a probabilistic layer, usually a normally distributed one with a mean of $\mu$ and a standard deviation of $\sigma$. During training/testing phase,  samples are drawn from this layer for prediction and also for aleotoric uncertainty quantification\footnote{This assumes the ML architecture chosen is able to give high accuracy before  replacing the output layer.}.  The problem with this approach is to  figure out how to learn the   parameters of this Normal distribution. This can be  solved by defining a new cost function, negative log-likelihood (NLL) that represents the loss between a distribution and the true output label. 

The NLL is equivalent to maximizing the likelihood of observing a data  given a distribution with its parameters.  In NLL, the logarithmic probabilities associated with each class is summed up  for a dataset. This closely resembles the cross entropy loss function except in cross entropy,  the last classification activation  is  implicitly applied before taking the logarithmic transformation,   while in NLL this is not the case. The NLL is given as, 

\begin{equation}\label{nll}
	\text{NLL} = - \log P(y_i|X_i;\mu, \sigma). 
\end{equation}

With NLL as a cost function and the last layer as a probabilistic layer, the aleotoric uncertainty can thus be modeled as described in Algorithm \ref{aleo}.  The independent normal layer can be implemented in any modern day ML packages. In our case, we used the TensorFlow Probability package \cite{dillon2017tensorflow} to model such layer.
\begin{algorithm}
	\caption{method to measure Aleotoric uncertainty }
	\begin{algorithmic}[1]\label{aleo}
		\renewcommand{\algorithmicrequire}{\textbf{Input: }}
		\renewcommand{\algorithmicensure}{\textbf{Output:}}
		\REQUIRE $\mathbb{D}(X_i,y_i)$, replace output layer with probabilistic node: $\mathcal{N}(\mu,\sigma)$, define optimizer as \textit{rmsProp} and set it's learning rate, set $num\_epochs$ for training.
		\ENSURE  $\hat{y_i}$ 
		\\ 
		\FOR {$epoch = 1$ to $num\_epochs$}
		\STATE i) Calculate loss and gradients using NLL $\eqref{nll}$ and the defined optimizer
		\STATE  ii) Apply gradients and update weights
		\STATE iii) Monitor loss and accuracy
		\ENDFOR
		\STATE Determine the parameters $\mu$ and $\sigma$ from the output layer
		\end{algorithmic} 
\end{algorithm}

Once the mean and standard deviation is determined, we can then easily figure out the 95\% confidence interval for the trained data or even for the test data.  For classification problems, the last layer can be  modeled as a categorical distribution, where for each class, there is a learned distribution  and based on the learned parameters for these distribution,  aleotoric uncertainty can be measured across classes. For more details on the implementation in TensorFlow, one can refer to \cite{dillon2017tensorflow}. 

Until now, we modeled the aleotoric uncertainty in a AI/ML model as a probabilistic layer at the output and a custom loss function as NLL, the next part is to model the epistemic uncertainty caused by the model itself. To include this epistemic uncertainty, the weights  associated with each layer  will now be considered as a random variable with a given probability distribution rather than a single deterministic value, which was the case before in a normal neural networks. The parameters of these weight distribution are then learnt by Bayes backpropagation algorithm, mention in detail in \cite{blundell2015weight}. In short, the difference between this type of Bayesian neural network and a normal neural network can  be summarized as,
\begin{itemize}
	\item Classic neural networks: the weights are, $\theta_i = \hat{\theta}_i$
	\item Probabilistic neural network or Bayesian Neural networks:  the weights are sampled from: $\theta_i \sim N(\hat{\mu}_i, \hat{\sigma}_i)$.
\end{itemize}

In the feed-forward pass, a sample from these weights is used to determine the output and then the Bayesian back prop is used to determine the distribution parameters.
Since, the weights are assumed to be a random variable, the first step is to determine their distribution. From Baye's theorem, the distribution of weights given the training data  $\mathbb{D}$  is given as,
\begin{equation} \label{bayes}
	\begin{split}
		p(\theta|\mathbb{D})&=\frac{p(\mathbb{D}|\theta)p(\theta)}{ \int p(\mathbb{D|\theta'}) p(\theta') d\theta'},\\
	\end{split}	
\end{equation} 
where $p(\theta)$ in the above equation is called the prior distribution, $p(\mathbb{D}|\theta)$ is the likelihood of observing data given weights,  $p(\theta|\mathbb{D})$ is  the posterior distribution and the denominator is the normalizing constant. In principle, the Bayesian learning works simply by,
\begin{itemize}
	\item Assume a prior distribution for weights, $p(\theta)$.
	\item Using training data $\mathbb{D}$ to determine the likelihood $p(\mathbb{D}|\theta)$
	\item  Finally determine the posterior density $p(\theta|\mathbb{D})$ using Bayes theorem.
\end{itemize}

This looks easier but the main challenge is in determining the   normalizing constant as it  involves solving or approximating a complicated solution to the integral.  One of the popular method to approximate it is with the variational Bayes, which is a very popular technique in the AI/ML domain\cite{kingma2013auto}.  Variational Bayes approximates the posterior distribution with a secondary function, known as variational posterior which is of a known form. This approximation   may lead to a posterior that  may be very inaccurate. This however can be overcomed by  tuning the function parameters so that it matches to the original posterior distribution as much as possible.  Let  $q(\theta|\phi)$  be the approximated posterior  distribution (variational posterior) instead of the true posterior density, $p(\theta|\mathbb{D})$,  parameterized by  $\phi$.  Thus, to approximate the variational posterior as close as possible to the original posterior distribution, the difference between  them should be as minimum as possible\cite{blundell2015weight,hinton1993keeping}.  This difference between two distributions  can be easily measured  by  Kullback-Leibler divergence (KLD). Therefore, the KLD between $q(\theta|\phi)$ and the true posterior $p(\theta|\mathbb{D})$ is given as,

\begin{equation}\label{dkl}
	\begin{split}
	\text{KLD} & (q(\theta | \phi)|| p(\theta | \mathbb{D} )) = \int  q(\theta | \phi) \log \bigg(
   \frac{q(\theta | \phi)}{p(\theta |  \mathbb{D} )} \bigg) d \theta \\
    &=  \int  q(\theta | \phi) \log \bigg(
    \frac{q(\theta | \phi) p(\mathbb{D} )  }{p(\mathbb{D} | \theta ) p(\theta)} \bigg) d \theta \\
    &= \int  q(\theta | \phi)  \log p(\mathbb{D} ) d \theta  + \int q(\theta | \phi) \log\bigg(\frac{q(\theta | \phi)}{p(\theta)}\bigg) d \theta \\
    & ~~~~~~~~~~~~~~~~~~~~~~~~~~~~~~~ -   \int  q(\theta | \phi)  \log p(\mathbb{D}  | \theta) d \theta 
\end{split}
\end{equation}
On further expanding the terms in \eqref{dkl}, the first term  will reduce to, $ \int  q(\theta | \phi)  \log p(\mathbb{D} ) d \theta =  \log p(\mathbb{D} )  \int  q(\theta | \phi)  d \theta = \log p(\mathbb{D} ) $. The second term is KLD($ q(\theta | \phi) ||  p(\theta)$), and the last term is nothing but expectation of the NLL of  $\log p(\mathbb{D}  | \theta) $ under the variational posterior $q(\theta | \phi)$.  Since the first term is constant, we can write \eqref{dkl} as a loss function- 	$L(q| \mathbb{D} ) $ as,

\begin{equation} \label{elbo}
	L(q| \mathbb{D} ) =  \text{KLD}(q(\theta | \phi) ||  p(\theta)) - \mathbb{E}_{q(\theta | \phi)} (\log p(\mathbb{D}  | \theta))
\end{equation}
Taking negative of the above equation \eqref{elbo}, will represent  the  lower bound on the log-evidence, and is known as the evidence lower bound (ELBO) loss as $\text{KLD}(q(\theta | \phi) ||  p(\theta)) $ is always positive.  Hence, the minimization of \eqref{elbo} will be a maximization of ELBO loss.

Maximizing this ELBO loss  represents a trade-off between the KLD term and expected log-likelihood term. On  one hand, the divergence between the variational posterior ($q(\theta | \phi)$) and  actual posterior ($p(\theta)$) should be as small as possible but on the other hand, the variational posterior parameters should maximize the expectation of the log-likelihood $\log p(\mathbb{D}  | \theta)$, implies that the model will assign a high likelihood to the data. Therefore, with this new ELBO loss, and weights modeled as  a random variable with a distribution, the epistemic uncertainty can be modeled as described in Algorithm \ref{epistemic}.
\begin{algorithm}[!ht]
	\caption{method to measure Epistemic uncertainty }
	\begin{algorithmic}[2]\label{epistemic}
		\renewcommand{\algorithmicrequire}{\textbf{Input: }}
		\renewcommand{\algorithmicensure}{\textbf{Output:}}
		\REQUIRE $\mathbb{D}(X_i,y_i)$,  define optimizer as \textit{rmsProp} and set its learning rate, set $num\_epochs$ for training.
		\ENSURE  $\hat{y_i}$ \\
		\textit{Initialization}\\
			\STATE   a) Assign a prior distribution with density $p(\theta)$ to weights $\theta$. This can be as simple as a normal Gaussian distribution.
		b) Assign the weights to variational posterior with density $q(\theta| \phi)$ with some trainable parameter $\phi$.\\
			\FOR {$epoch = 1$ to $num\_epochs$}
		\STATE i) Calculate loss and gradients using NLL part from $\eqref{elbo}$ and optimizer\\
		 ii) learn $\phi$ using KLD from \eqref{elbo} to approximate the variational posterior as close to original posterior.\\
		  iii) Apply gradients and learn weight parameters\\
		 iv) Monitor loss and accuracy\\
		\ENDFOR
		\STATE Determine class probabilities or mean/variance.
	\end{algorithmic} 
\end{algorithm}
 \begin{figure*}[t]
	\begin{minipage}[t]{0.45\linewidth}
		\includegraphics[width=\linewidth]{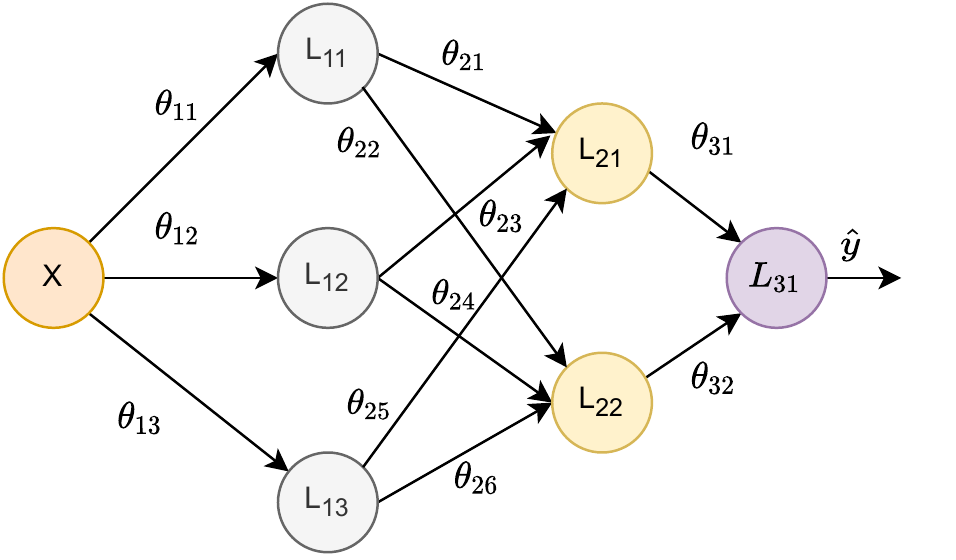}
		\subcaption{Normal ML model with deterministic weights $\theta_{ij}$, where $i,j$ depend on the layer to layer connection.}
	\end{minipage}%
	\hfill%
	\begin{minipage}[t]{0.45\linewidth}
		\includegraphics[width=\linewidth]{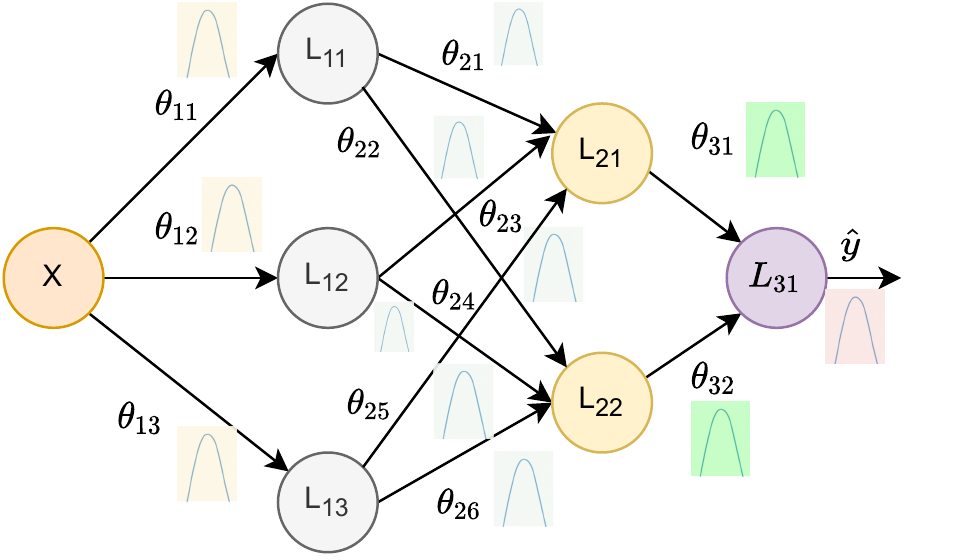}
		\subcaption{Probabilistic neural network where weights and output layer are represented as a distribution.}
	\end{minipage} 
	\caption{A two layer ML model showing the difference between a normal ML model and a probabilistic ML model.}
	\label{probModel}
\end{figure*}

Now, to combine both aleotoric and epistemic uncertainty in a AI/ML model, we can just combine these layers with weights as random variable with a probabilistic output layer as described before. This type of a model as compared to a conventional AI/ML model can be easily visualized as shown in Fig. \ref{probModel}. In the next section, we will look into a real example of modeling uncertainty in a WiFI CSI sensing application.

%

\section{WiFi CSI Sensing} \label{wifiSensing}

The  CSI from a WiFi device provides amplitude and phase information that can be used in applications such as motion/no-motion detection. Since WiFi is ubiquitous,  deploying any CSI application is easy and very cost effective.  In this section, we will look at one of the application of WiFi sensing for motion/no-motion detection. Mathematically, the measured baseband to baseband CSI in a WiFi is given as\cite{ma2019wifi,tadayon2019decimeter},
\begin{equation}\label{csi}
	\begin{split}
		H_{i,j,k}& =\mathlarger{‎‎\sum}_{n=1}^{N} a_n \Phi_{i,j} e^{-j2\pi f_k [d_{i,j,n}/c+ \tau_i+  \nabla_t +  \eta\nabla_f]},\\
		H_{i,j,k} & \in \mathbb{C}^{N_{Tx}\times N_{Rx}\times N_{sc} \times N_{samples} },
	\end{split}
\end{equation}
where, $a_n$ is the amplitude of the received signal, $ \Phi_{i,j}$ is the beamforming matrix, $f_k$ is the $k^{th}$ carrier frequency, $\tau_i$ is the time delay from Cyclic Shift Diversity (CSD) of the $i^{th}$ transmit antenna, $\nabla_t$ is the Sampling Time Offset (STO), $\eta$ is the Sampling Frequency Offset, and $\nabla_f=f_k'/f_k-1$. Therefore, the WiFi CSI data represents a 4D (dimensional) complex vector as shown in Eq. \eqref{csi} with $(N_{Tx}, N_{Rx}, N_{sc}, N_{samples})$ as its dimensions, where  $N_{Tx}$ is number of transmit antennas, $N_{Rx}$ is the number of received antennas, $N_{sc}$ is the number of subcarriers, and $N_{samples}$ are the number of collected time samples.
The CSI sampling rate in our experiment is fixed at 100 Hz, which is enough to capture any human motion signatures\cite{li2017indotrack}. Also, we will utilize CSI amplitude to generate input features for our  probabilistic AI/ML model\footnote{Phase information is also useful but is very sensitive to environment changes and can lead to lot of false-positive cases.}.

\begin{figure*}[!t]
	\begin{minipage}[t]{0.45\linewidth}
		\includegraphics[width=\linewidth]{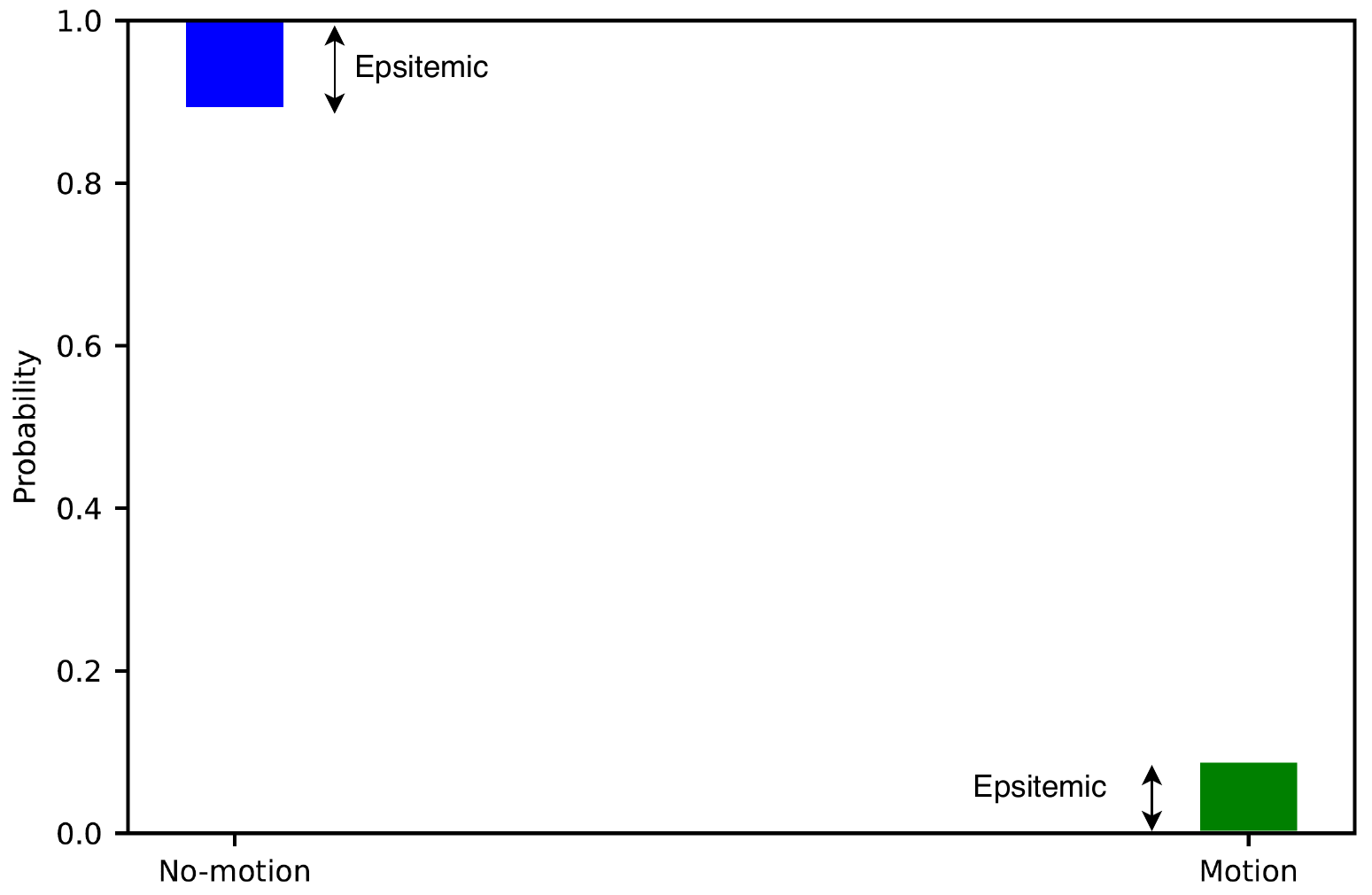}
		\subcaption{An example from  test Home-1, where the model is very confident  about its prediction (less aleotoric and less epistemic uncertainty). The predicted and true label are both `No-motion'.}
	\end{minipage}%
	\hfill%
	\begin{minipage}[t]{0.45\linewidth}
		\includegraphics[width=\linewidth]{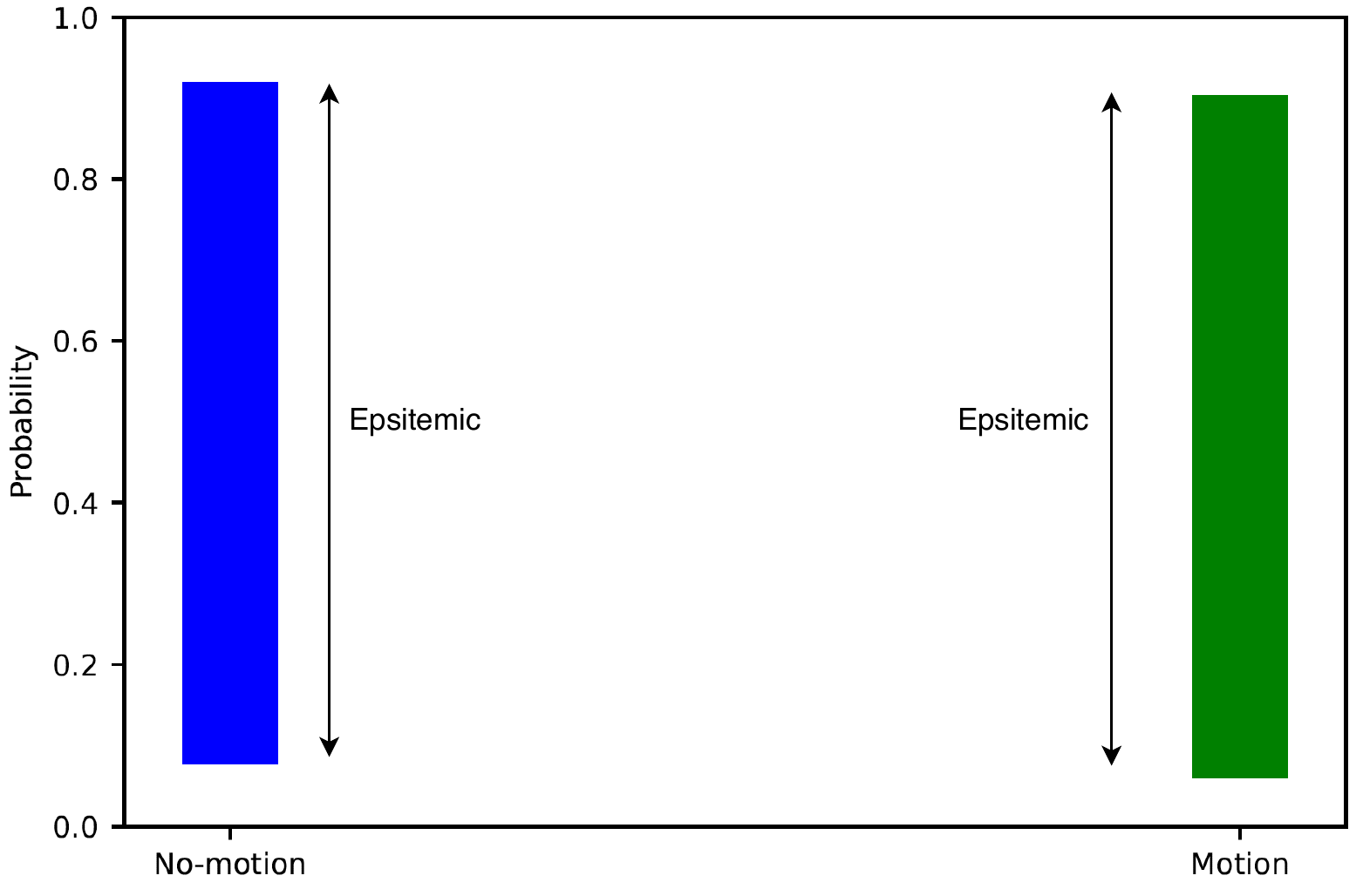}
		\subcaption{An example from test Home-3, where the model is very uncertain about its prediction (high aleotoric and high epistemic uncertainty). The prediction  and  true label are both `Motion'.}
	\end{minipage} 
	\caption{Aleotoric (different class probability) and Epistemic uncertainty in different homes.}
	\label{quant}
\end{figure*}

The intuition behind using CSI amplitude for motion/no-motion detection is that the human motion  causes amplitude fluctuations, which can discriminate motion and no-motion case\cite{ma2019wifi,liu2020human}. In our experiment, we will  consider 4 different homes, where we collect CSI data and label it as motion and no-motion depending on the subject is moving or not-moving there. 
To observe the uncertainty in the predictions of our model, we would use permutations of 3 homes for training and  keep one home for testing. Intuitively, this will represent an interesting scenario where the test data will behave mostly as a out of distribution data that the model will not have  learned during training. Since the data is limited, we have to rely on generating features from the CSI data. 

To begin with, we will pre-process the raw CSI data collected from these 4 homes so as to remove outliers and noise in the CSI data. These steps will also mitigate some of the hardware ailments to some extent that may affect CSI data quality adversely\cite{ma2019wifi,liu2020human}. The brief details of each step are as follows,
\begin{itemize}
	\item\textit{Hampel filtering}: Hampel filter  removes and replaces any outlier points or jitter in the CSI data by the median value of the hampel filter window. 
	\item \textit{Subcarrier selection}: Choosing the right subcarriers is  very important as different subcarriers are affected differently in an environment  and a particular set of subcarriers  can't be generalized for every home\cite{kaltiokallio2012enhancing}. We rank subcarriers in terms of having high mean power and variance and choose top 5 subcarriers from that ranking list. The idea is that motion causes the amplitude fluctuations with a high mean power, while the noisy subcarrier will have high variance but not high mean power\footnote{Since our main aim is to quantify uncertainty, any other subcarrier selection method can be used here. For simplicity, we will use the mean power and variance as a metric to choose subcarriers.}.
	\item \textit{Standardize and mean of top 5 subcarriers}: After selecting the top 5 subcarriers, random noise across subcarriers can be  further reduced by taking the mean across these subcarriers. Apart from removing noise, it simplifies the output as it   creates a single 1D time series.
	\item \textit{Feature engineering}: Before feeding the pre-processed 1D time series data to ML algorithm, we will generate 7 high level features\footnote{More features can be used for analysis but since our main aim was to determine uncertainty, we kept it to the 7 features.}- a combination of both temporal and spectral features. The temporal features chosen are  sample entropy, skewness, and kurtosis while the selected spectral features are binned entropy, Fourier entropy, maximum Doppler and doppler spread. 
\end{itemize}
Finally, after all these  steps, we will end up with a 1D time series of shape, $N_{samples} \times 7$ as our input and the output would be $N_{samples} \times 2$, where the last dimension in it refers to the two classes of motion and no-motion.  In the next section, we will discuss more on the model and corresponding results on uncertainty quantification in these 4 homes.

\section{ML Modeling and  Results} \label{QuantResults}

For our experiments, we will create a very simple two  probabilistic model layers  with the first layer having 4 nodes, and second one with 2 nodes. The output layer is modeled as a probabilistic layer to capture the aleotoric uncertainty.  The prior  distribution  and trainable posterior distribution in the first layer are both set as a  multivariate Normal distribution with activation set to rectified linear unit (\textit{relu}). The second layer has 2 dense  nodes  with no activation, while the prior and posterior distribution of this layer is also set to a multivariate Normal distribution\footnote{One can define any distribution but for simplicity and tractability, we kept it multivariate Normal distribution.}. The final layer  in this model is defined as a one hot categorical layer with 2 outputs each representing a distribution. Hence, the output will have four outputs representing  means and standard deviations (4 outputs) for each class. The  divergence function in both layers can be defined  empirically. However,  this may not be required if Normal distribution is  assumed as prior and posterior distribution as it has a tractable analytical solution for the divergence. 

The model is then trained on 3-homes for 200 epochs while testing on one left out home. The batch size is kept to 4, and optimizer used to train the model is \textit{RMSprop}.
\setlength{\extrarowheight}{1pt}
\begin{table}[t]
	\centering	
	\caption{Test home results with accuracy and mean entropy  for no-motion and motion class.}
	\begin{tabularx}{\columnwidth}{|X|X|X|X|}
		\hline
		\multicolumn{1}{|l|}{Test home} & Accuracy (\%) & $\bar{H}$(No-motion) & $\bar{H}$(Motion)	 \\ \hline
		Home-1 & 80.55 & 0.43 & 0.83  \\ \hline
		Home-2 & 76.78 & 0.71 & 0.71   \\ \hline
		Home-3 & 57.50 &0.45 & 0.45  \\ \hline
		Home-4 & 82.03 & 0.12 & 0.41  \\ \hline
	\end{tabularx} 
	\label{table1}
\end{table}
This model is implemented in TensorFlow with TensorFlow's probability module with layers such as  \textit{DenseReparameterization} and \textit{OneHotCategorical}\cite{dillon2017tensorflow}.  Figure \ref{quant} shows one example  from two test homes-1 and -3, while being trained on the rest 3 homes. As one can observe, the model  trained on homes-2, 3, and 4, is  very certain  on his prediction in that one example data  from home-1,  while the model  trained on homes-1,2, and 4,  is very uncertain in his prediction in that one  example data from home-3.  Although, one may observe that in both examples, the prediction match very well with the true label but with probabilistic modeling, the model was able to convey the most important information that is how much uncertain or confident the model was in its prediction for that example data.  

\begin{figure}[t]
	\centering
	\includegraphics[width=1\linewidth]{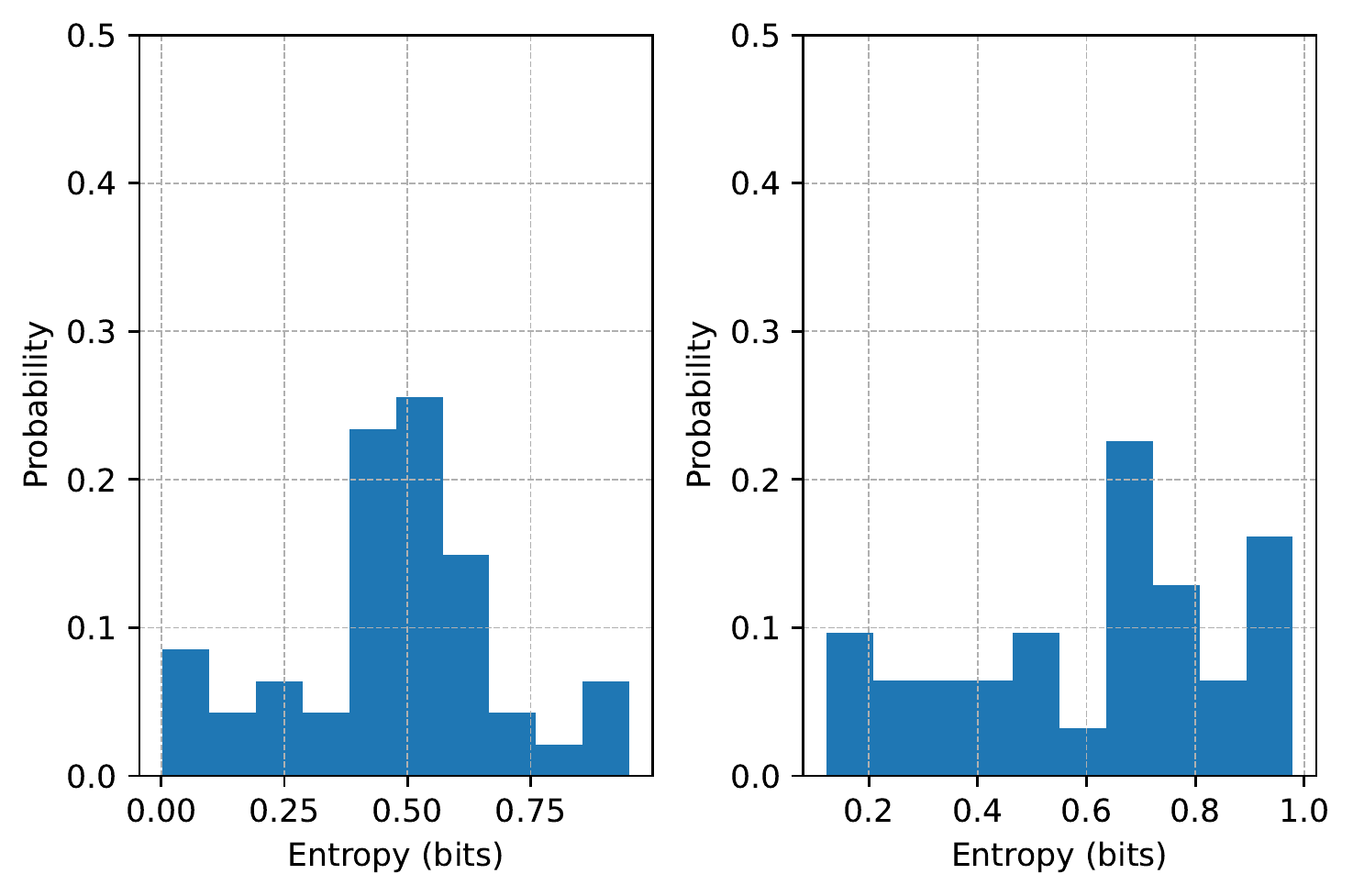}
	\caption{Entropy measure for test home 3 for each class, where the model is trained on Home-1,2 and 4. This test home represents the highest uncertainty for both of it's classes. }
	\label{home-3}
\end{figure}

To analyze the model's uncertainty across the full test set,  we can use  entropy of the distribution as a metric for uncertainty.  The entropy is given as, $H_i = - \sum_{j=1}^{n_t}p_i \log_2(p_i)$, where $n_t$ is the number of samples of a class and $i$ represents  one of those samples. Thus, higher the entropy of a class, higher is the uncertainty. Table \ref{table1} presents the mean entropy for each class in all the test home cases, and Fig. \ref{home-3}  shows the Entropy for each class in case of test home-3.  
There are few important observation from these results,
\begin{itemize}
	\item Although the accuracy for test home-1 is 80\% but the mean entropy for motion class is very high as compared to the no-motion class. This implies that the model is very uncertain in its prediction on motion classes as compared to the no-motion class. Same can be inferred for  test home-4.
	\item Test home-2 and home-3 are  having equal mean entropies across classes but the accuracy is lower in home-3, which means the model is performing really bad in terms of predictions for both  the classes and is very uncertain in its predictions as compared to other homes especially for no-motion class. 
	\item The best performance of the model is on test home-4 in the no-motion class.
\end{itemize}
Finally, with this probabilistic modeling example on a WiFi CSI data for uncertainty quantification, one can easily understand the importance of such models in wireless sensing domain, where data is highly dependent on the measurement environment and prone to errors. This ability to convey the confidence or uncertainty with a model's prediction is therefore very handy.  On the contrary, there are some disadvantages to this probabilistic ML modeling, 1) training  of probabilistic machine learning models is computationally expensive and takes a lot of time to converge, 2) A reliable prior distribution and the assumed variational posterior is a pre-requisite as can be seen from previous sections, and finally 3) the implementation can be challenging if there are memory or computation constraints. 

\section{Conclusion and Future Work} \label{ConclusionFuture}
In this paper, we describe the need of uncertainty quantification of  AI/ML models and laid the foundations for modeling the two types of uncertainty (aleotoric and epistemic uncertainty) in a AI/ML model.  We also describe the different cost functions and Bayesian technique used in the quantification of  aleotoric and epistemic uncertainty. As an example, we looked into a wireless sensing case of motion/no-motion detection with WiFi CSI and quantified uncertainty for each test home where the model was trained on other homes. Furthermore, to quantify uncertainty on the whole test set, we used Entropy as a broader metric.  The results highlight the need of uncertainty quantification, where we observed that even though in test Home-1,  the prediction accuracy was high  but the predictions for motion class as compared to no-motion class were very uncertain. We also saw the case of test Home-3 having high epistemic as well as high aleotoric uncertainty with less prediction accuracy.  These results are valuable as it forces one to rethink on the data collection strategy or feature engineering to handle aleotoric uncertainty or the need to collect more data or iterate over different model architecture to overcome epistemic uncertainty. In future, we would explore different methods to overcome the aleotoric and epistemic  uncertainty in wireless sensing applications.


%

\section*{Acknowledgment}

The authors would like to thank Morris Hsu, Maxim Arap, Ravi Ichapurapu, Swamy Inti,  and Abhishek  Sanaka for their great support and help during this work.

\bibliography{biblography}

\end{document}